\documentclass[10pt,a4paper]{article}
\usepackage{amsmath,amssymb,amsfonts}
\usepackage{graphicx}
\usepackage{hyperref}
\usepackage[margin=0.75in]{geometry}
\usepackage{float}
\usepackage{bm}
\usepackage{booktabs}
\usepackage{tabularx}

\usepackage{setspace}
\usepackage{titlesec}
\titlespacing*{\section}{0pt}{8pt}{4pt}
\titlespacing*{\subsection}{0pt}{6pt}{3pt}
\titlespacing*{\subsubsection}{0pt}{4pt}{2pt}

\AtBeginDocument{
  \setlength{\abovedisplayskip}{4pt}
  \setlength{\belowdisplayskip}{4pt}
  \setlength{\abovedisplayshortskip}{2pt}
  \setlength{\belowdisplayshortskip}{2pt}
}

\let\oldbibliography\thebibliography
\renewcommand{\thebibliography}[1]{%
  \oldbibliography{#1}%
  \setlength{\itemsep}{0pt}%
  \setlength{\parskip}{0pt}%
}

\title{A Reduced-Order Dynamical Model for the Ignition of
Diver-Induced Cohesive Silt-Out on Sloping Beds}

\author{
Sandy H. S. Herho$^{1,2,3,*}$,
Iwan P. Anwar$^{3,4}$, Umar Abdurrahman$^{5}$, Faruq Khadami$^{3,4}$,\\
Alfita P. Handayani$^{1,6}$, Karina A. Sujatmiko$^{4}$,
and Dasapta E. Irawan$^{2}$
}

\date{}

\def\Address{$^{1}$Center for Agrarian Studies, Bandung Institute of Technology,
Bandung, West Java 40132, Indonesia\\
$^{2}$Applied Geology Research Group, Bandung Institute of
Technology, Bandung, West Java 40132, Indonesia\\
$^{3}$Indonesian Subaquatic Sport Association (POSSI), North Jakarta,
DKI Jakarta 14240, Indonesia\\
$^{4}$Applied and Environmental Oceanography Research Group, Bandung Institute of Technology, Bandung, West Java 40132, Indonesia\\
$^{5}$Faculty of Fisheries and Marine Science, Padjadjaran University, Sumedang, West Java 40600, Indonesia\\
$^{5}$Spatial Systems and Cadaster Research Group, Bandung Institute of Technology, Bandung, West Java 40132, Indonesia}

\def\corrAuthor{Corresponding Author}
\def\corrEmail{sandyherho@itb.ac.id}

\begin{document}
\maketitle

\begin{center}
\small
\Address\\[2pt]
$^{*}$\corrAuthor: \href{mailto:\corrEmail}{\corrEmail}
\end{center}

\begin{abstract}
\noindent
A downward fin thrust near a cohesive seabed lofts sediment. Whether this cloud settles or organizes into a self-sustaining down-slope current determines if a diver loses visibility briefly or catastrophically. We reduce the layer-averaged balances of fluid mass, sediment mass, and momentum to a temporal slab of fixed thickness in the weakly entraining limit. This yields an autonomous planar vector field in dimensionless near-bed speed and suspended load, governed by three dimensionless groups: drag against settling, erosion strength, and near-bed concentration. The field possesses a quiescent equilibrium at the origin and, above a critical bed slope obtained in closed form as the ratio of near-bed concentration times drag-settling number to erosion-strength number, an interior saddle equilibrium. The saddle's stable manifold partitions the state space into basins of decay and unbounded growth, establishing the exact threshold between self-limiting and igniting disturbances. On a flat bed, the origin attracts globally for all admissible erosion laws. For representative silt, the ignition threshold is near a 17-degree bed angle. The closed-form structure is confirmed numerically to machine precision. A closure study demonstrates the critical slope is robust in existence but closure-dependent in value, identifying the linear erosion law as the conservative choice. The model thus isolates bed slope as the controlling parameter for ignition.
\end{abstract}


\section{Introduction}

The sudden loss of visibility that follows disturbance of a fine-grained bottom,
known among divers as a silt-out, is a recurrent hazard of underwater work. A
single downward thrust of a fin near a cohesive seabed resuspends a cloud of
fine sediment that can reduce the visual range to a few centimeters within
seconds, with immediate consequences for orientation, gas management, and safe
ascent. The governing question is not whether a disturbance lifts sediment,
which it plainly does, but whether the lofted cloud settles back to the bed once
the forcing stops or instead organizes into a persistent, self-feeding
down-slope flow. The two outcomes differ by orders of magnitude in duration and
extent, and the distinction between them is a dynamical one: it is the same
decay-against-runaway alternative that the sediment-transport literature studies
under the name of autosuspension.

A sediment-laden current descending a slope can maintain its own suspension when
the potential energy released by the settling load supplies the turbulent energy
required to keep the grains aloft, which is the energetic criterion for
autosuspension \cite{bagnold1962}. Phase-plane analysis of the coupled evolution
of the current speed and its effective density separates the trajectories that
decay from those that run away and furnishes a criterion for self-sustaining
motion \cite{pantin1979}. An erosive gravity current possesses an unstable
ignitive equilibrium below which disturbances die out and above which they
accelerate and entrain sediment toward a distinct, stable catastrophic state, so
that the ignitive equilibrium itself defines the threshold for a self-sustaining
flow \cite{parker1982}. These results rest on layer-averaged balances of fluid
mass, sediment mass, and mean momentum for a turbid underflow
\cite{pfp1986,meiburg2010}, closed by empirical laws for the exchange of sediment
with the bed for both noncohesive and cohesive material
\cite{partheniades1965,garciaparker1991}. The autosuspension threshold is
supported experimentally, though its precise location is sensitive to the erosion
law adopted \cite{southard1981}.

The diver-induced silt-out lies within this framework, since the resuspended
cloud on a sloping bottom is a small, transient gravity current
\cite{simpson1982,meiburg2010} whose fate is governed by the same competition
between slope-driven acceleration and settling that controls the ignition of
turbidity currents. What the existing treatments do not supply is a minimal model
at the scale of a diver disturbance that connects the resuspension event to the
ignition criterion without the apparatus of a spatially resolved field solver
\cite{meiburg2015}, an apparatus whose turbulence closure is difficult to defend
at diver-scale Reynolds numbers, whose output is a set of fields rather than a
transparent statement about the existence of a threshold, and whose
layer-averaged reductions themselves remain under active scrutiny
\cite{hu2015}. A complementary line of work has instead built compact,
single-purpose idealized solvers whose numerical core is legible and whose
analysis emphasizes verified structure over calibrated prediction, an approach
that has been applied to shear-driven instability \cite{herho2025kh2d},
collective animal motion \cite{herho2026dewikadita}, nonlinear dispersive waves
\cite{irawan2026sangkuriang}, and wave attenuation in coastal vegetation
\cite{herho2026waveatten}, and in which dynamical-systems and
information-theoretic diagnostics repeatedly expose structure that first-moment
descriptions leave hidden. A reduced-order treatment of diver-induced silt-out
belongs to that line, because the outcome of interest is precisely the
qualitative one, decay against ignition, that a low-dimensional dynamical system
is designed to resolve.

This paper develops and analyzes such a model. The layer-averaged balances of
fluid mass, sediment mass, and momentum for a slab of resuspended cloud are
reduced, under a small set of explicitly stated and justified assumptions, to an
autonomous planar vector field for a dimensionless near-bed speed and a
dimensionless suspended load, governed by three dimensionless groups with the bed
slope as the control parameter. The equilibria of that field, their linear
stability, the critical slope at which an ignitive equilibrium appears, the
degenerate structure of the quiescent state, the global attraction of the flat
bed, and the invariant manifold that separates decay from ignition are all
obtained in closed form. The closed-form structure is then confirmed numerically,
its qualitative content is shown to be robust to the erosion closure, and its
physical reading is developed with explicit attention to the limitations that
bound its scope. The governing equations and their reduction are derived first,
the numerical and diagnostic methods are described next, the computed structure
is reported in full, and the results are interpreted and delimited in the
discussion.

\section{Methods}
\subsection{Model Description}

\begin{figure}[H]
\centering
\includegraphics[width=0.86\linewidth]{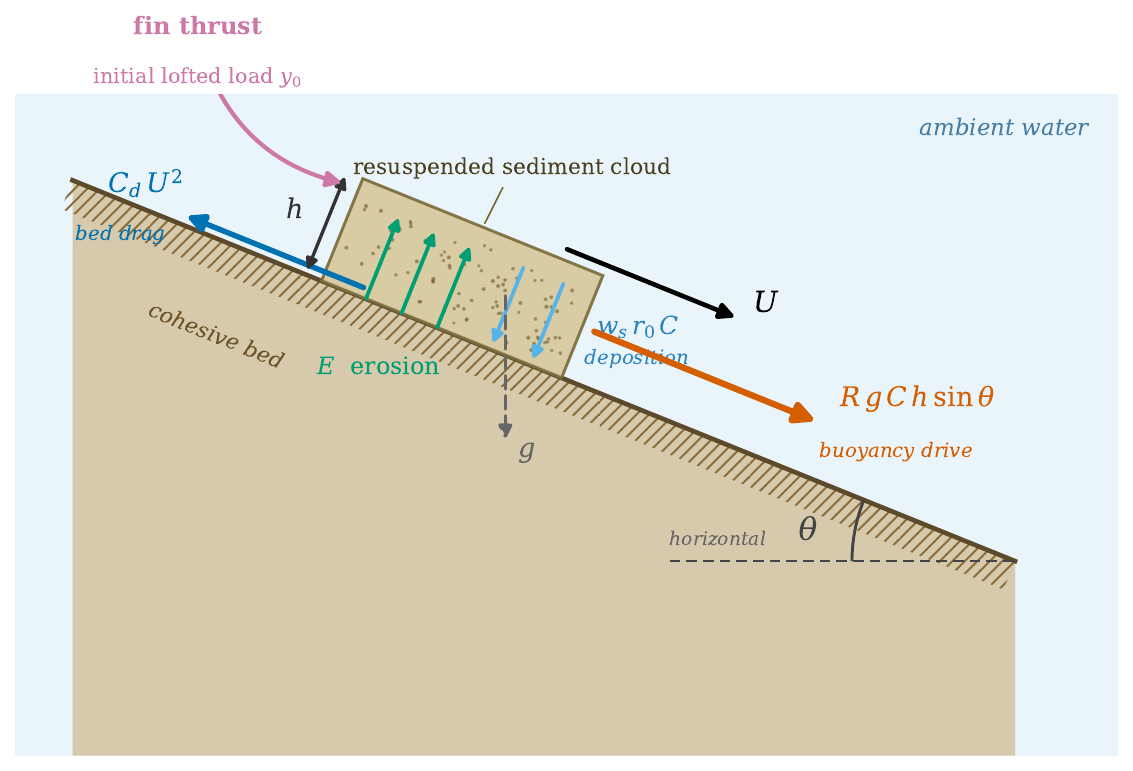}
\caption{Physical configuration of the diver-induced silt-out. A fin thrust
lofts fine cohesive sediment into a near-bed slab of thickness $h$ that moves
down a bed inclined at angle $\theta$. The down-slope component of the reduced
gravity, $R\,g\,C\,h\sin\theta$, drives the slab; the bed drag $C_d U^2$ retards
it; the slab entrains sediment from the bed through the erosion flux $E$ and
returns it through the deposition flux $w_s r_0 C$. The initial lofted load $y_0$
set by the thrust is the disturbance whose fate, decay or ignition, the model
determines.}
\label{fig:schematic}
\end{figure}

The physical setting is a diver-scale disturbance of a cohesive bottom inclined
at an angle $\theta$ to the horizontal. A downward fin thrust resuspends a volume
of fine sediment into a near-bed cloud; once the thrust ceases, the cloud is left
to evolve under gravity, drag, settling, and its own capacity to entrain further
sediment from the bed. The cloud is represented as a single layer-averaged slab
of sediment-laden fluid moving along the bed, and its evolution is traced in time
from the instant the external forcing stops. Figure~\ref{fig:schematic} shows the
physical configuration and the agencies acting on the slab. The derivation
proceeds from the integral balances for the slab, introduces the closures that
render the balances a closed system, reduces the system to two state variables
through a justified thickness assumption, nondimensionalizes it, and analyzes the
resulting planar vector field in closed form.

Let $U(t) \geq 0$ denote the layer-averaged down-slope speed of the slab at time
$t$, let $h(t) > 0$ denote its thickness, and let $C(t) \in [0,1)$ denote its
layer-averaged volumetric sediment concentration, each defined per unit area of
bed. Let $\rho_0$ be the density of the clear ambient water, $\rho_s$ the density
of the sediment grains, and $g$ the gravitational acceleration. A suspension of
concentration $C$ carries an excess density $C(\rho_s - \rho_0)$ over the ambient
fluid, and the buoyant acceleration it experiences along the bed, referred to the
ambient density under the Boussinesq approximation, is the reduced gravity
\begin{equation}
g' \;=\; g\,\frac{\rho_s - \rho_0}{\rho_0}\,C \;=\; R\,g\,C,
\qquad
R \;\equiv\; \frac{\rho_s - \rho_0}{\rho_0},
\label{eq:reduced_gravity}
\end{equation}
where $R$ is the submerged specific gravity of the sediment, close to $1.6$ for
quartz grains in seawater. The three integral balances for the slab, per unit bed
area, express conservation of ambient fluid, of suspended sediment, and of
down-slope momentum, and follow the layer-averaged description of buoyancy-driven
sediment flows \cite{simpson1982,meiburg2010}. Ambient water is entrained across
the upper interface at a rate proportional to the speed through a dimensionless
entrainment coefficient $e_w$, which thickens the layer,
\begin{equation}
\frac{\mathrm{d}h}{\mathrm{d}t} \;=\; e_w\,U .
\label{eq:fluid_mass}
\end{equation}
The suspended sediment per unit bed area, the load $q \equiv C h$, changes through
the difference between erosion of the bed and deposition onto it, both scaled by
the settling velocity $w_s$,
\begin{equation}
\frac{\mathrm{d}}{\mathrm{d}t}\!\left( C h \right)
\;=\; w_s\,\big( E - r_0\,C \big),
\label{eq:sediment_mass}
\end{equation}
where $E$ is a dimensionless entrainment function of the bed and $r_0 C$ is the
near-bed concentration that feeds deposition, $r_0$ being the ratio of the
near-bed to the layer-averaged concentration. The down-slope momentum per unit
bed area, $\rho_0 h U$ under the Boussinesq approximation, is driven by the
down-slope component of the buoyancy force and retarded by the bed shear stress
$\tau_b$,
\begin{equation}
\rho_0\,\frac{\mathrm{d}}{\mathrm{d}t}\!\left( h\,U \right)
\;=\; \rho_0\,R\,g\,C\,h\,\sin\theta \;-\; \tau_b .
\label{eq:momentum_full}
\end{equation}
Equations~\eqref{eq:fluid_mass}--\eqref{eq:momentum_full} are the temporal
layer-averaged balances for the slab; they are the box-model counterpart of the
spatially developing three-equation model of a turbid underflow, whose
formulation and range of validity are established elsewhere
\cite{pfp1986,hu2015}.

Three closures render the balances a closed system. The bed shear stress is taken
quadratic in the layer speed through a single drag coefficient $C_d$, which
identifies the skin friction that mobilizes sediment with the form drag that
retards the layer,
\begin{equation}
\tau_b \;=\; \rho_0\,C_d\,U^2 .
\label{eq:bed_shear}
\end{equation}
Combining~\eqref{eq:bed_shear} with a critical shear stress $\tau_c$ below which
the bed does not erode defines a natural speed scale, the erosion-threshold speed
$U_c$, at which the bed shear equals the critical stress,
\begin{equation}
\frac{\tau_b}{\tau_c}
\;=\; \frac{\rho_0\,C_d\,U^2}{\tau_c}
\;=\; \left(\frac{U}{U_c}\right)^{\!2},
\qquad
U_c \;\equiv\; \sqrt{\frac{\tau_c}{\rho_0\,C_d}} ,
\label{eq:threshold_speed}
\end{equation}
so that the shear ratio governing erosion is exactly the square of the speed
measured in units of $U_c$. Erosion follows an excess-shear law of Partheniades
type, in which sediment is entrained only when the bed shear exceeds the critical
stress and at a rate proportional to the fractional excess
\cite{partheniades1965,garciaparker1991},
\begin{equation}
E \;=\; \alpha\,\max\!\left(\frac{\tau_b}{\tau_c} - 1,\; 0\right)
\;=\; \alpha\,\max\!\left( \Big(\frac{U}{U_c}\Big)^{2} - 1,\; 0 \right),
\label{eq:erosion}
\end{equation}
where the ramp $\max(\cdot,0)$ enforces that no erosion occurs below threshold and
the dimensionless erosion magnitude $\alpha$ is fixed by the Partheniades
erosion-rate parameter $M$, of dimension eroded mass per unit bed area per unit
time, through $\alpha = M/(\rho_s w_s)$, so that $w_s\alpha$ is the volumetric
entrainment rate at unit fractional excess \cite{partheniades1965}. The
entrainment coefficient in the fluid balance closes against the bulk Richardson
number $\mathrm{Ri} = R\,g\,C\,h/U^2$, with $e_w$ a decreasing function of
$\mathrm{Ri}$ that is largest for weakly stratified, energetic currents and
smallest for the strongly stratified, slow currents of interest here
\cite{pfp1986}.

A diver-induced cloud is dense and slow: its near-bed concentration is high, its
speed is at most comparable to the erosion-threshold speed $U_c$ of order
$10^{-1}\,\mathrm{m\,s^{-1}}$, and its bulk Richardson number is therefore large,
so that the entrainment coefficient $e_w(\mathrm{Ri})$ is small. In the strongly
stratified limit $e_w \to 0$, equation~\eqref{eq:fluid_mass} gives
$\mathrm{d}h/\mathrm{d}t = 0$, and the thickness remains at the value $h_0$ set by
the initial disturbance,
\begin{equation}
h(t) \;\equiv\; h_0 .
\label{eq:slab_assumption}
\end{equation}
This is the assumption that makes the model minimal: it removes the thickness as a
dynamical degree of freedom and retains the two variables, speed and load, whose
competition controls ignition. Its physical content is that the cloud entrains
ambient water slowly compared with the timescales on which its speed and load
evolve, which is the appropriate limit for a slow, dense, near-bed layer and is
conservative in the sense established below. Under~\eqref{eq:slab_assumption} the
load is $q = C h_0$, and dividing the momentum balance~\eqref{eq:momentum_full} by
$\rho_0$ while using~\eqref{eq:bed_shear} gives
\begin{equation}
h_0\,\frac{\mathrm{d}U}{\mathrm{d}t}
\;=\; R\,g\,\sin\theta\,q \;-\; C_d\,U^2,
\label{eq:momentum_slab}
\end{equation}
while the sediment balance~\eqref{eq:sediment_mass}, using~\eqref{eq:erosion}
and~\eqref{eq:threshold_speed}, becomes
\begin{equation}
\frac{\mathrm{d}q}{\mathrm{d}t}
\;=\; w_s\,\alpha\,\max\!\left( \Big(\frac{U}{U_c}\Big)^{2} - 1,\; 0 \right)
\;-\; w_s\,r_0\,\frac{q}{h_0} .
\label{eq:sediment_slab}
\end{equation}
Equations~\eqref{eq:momentum_slab} and~\eqref{eq:sediment_slab} are a closed pair
of autonomous ordinary differential equations for $U$ and $q$.

The pair is made dimensionless by referring the speed to $U_c$, the load to a
scale $q_c$ fixed below, and the time to the settling timescale of the cloud.
Introduce
\begin{equation}
x \;\equiv\; \frac{U}{U_c},
\qquad
y \;\equiv\; \frac{q}{q_c},
\qquad
s \;\equiv\; \frac{t}{t_{\mathrm{settle}}},
\qquad
t_{\mathrm{settle}} \;\equiv\; \frac{h_0}{w_s},
\label{eq:nondim_vars}
\end{equation}
in which $x$ is a scalar speed rather than a velocity, since the slab moves down
the slope and only the magnitude enters the balances, and $t_{\mathrm{settle}}$
is the time for the settling velocity to clear a cloud of thickness $h_0$. With
$U = U_c x$, $q = q_c y$, and $\mathrm{d}/\mathrm{d}t =
(w_s/h_0)\,\mathrm{d}/\mathrm{d}s$, the momentum balance~\eqref{eq:momentum_slab}
becomes
\begin{equation}
h_0\,\frac{U_c\,w_s}{h_0}\,\frac{\mathrm{d}x}{\mathrm{d}s}
\;=\; R\,g\,\sin\theta\,q_c\,y \;-\; C_d\,U_c^2\,x^2,
\label{eq:momentum_expand}
\end{equation}
and dividing through by $U_c w_s$ gives
\begin{equation}
\frac{\mathrm{d}x}{\mathrm{d}s}
\;=\; \frac{R\,g\,\sin\theta\,q_c}{U_c\,w_s}\,y
\;-\; \frac{C_d\,U_c}{w_s}\,x^2 .
\label{eq:momentum_pre}
\end{equation}
The load scale is fixed by requiring the buoyancy coefficient
in~\eqref{eq:momentum_pre} to reduce to the bed slope alone, which selects
\begin{equation}
q_c \;\equiv\; \frac{U_c\,w_s}{R\,g}
\qquad\Longrightarrow\qquad
\frac{R\,g\,\sin\theta\,q_c}{U_c\,w_s} \;=\; \sin\theta \;\equiv\; S,
\label{eq:load_scale}
\end{equation}
so that the momentum balance takes its final form
\begin{equation}
\;\frac{\mathrm{d}x}{\mathrm{d}s} \;=\; S\,y \;-\; P\,x^2\;,
\qquad
P \;\equiv\; \frac{C_d\,U_c}{w_s},
\label{eq:momentum_nondim}
\end{equation}
where $S$ is the dimensionless bed slope and $P$ is the drag-settling number, the
ratio of the bed drag at threshold speed to the settling velocity. Applying the
same substitution to the sediment balance~\eqref{eq:sediment_slab}, and dividing
through by $q_c w_s/h_0$, gives
\begin{equation}
\frac{\mathrm{d}y}{\mathrm{d}s}
\;=\; \frac{h_0}{q_c}\,\alpha\,\max\!\left( x^2 - 1,\; 0 \right)
\;-\; r_0\,y,
\label{eq:sediment_pre}
\end{equation}
and substituting the load scale~\eqref{eq:load_scale} into the coefficient
$h_0\alpha/q_c$ yields
\begin{equation}
\;\frac{\mathrm{d}y}{\mathrm{d}s} \;=\; A\,\max\!\left( x^2 - 1,\; 0 \right)
\;-\; r_0\,y\;,
\qquad
A \;\equiv\; \frac{\alpha\,h_0\,R\,g}{w_s\,U_c},
\label{eq:sediment_nondim}
\end{equation}
where $A$ is the erosion-strength number, the capacity of the bed to feed the
suspension, and $r_0$ retains its meaning as the near-bed concentration ratio.
Equations~\eqref{eq:momentum_nondim} and~\eqref{eq:sediment_nondim} are the
reduced model: an autonomous planar vector field
$\mathbf{f}(x,y) = (S y - P x^2,\; A\max(x^2-1,0) - r_0 y)$ on the state $(x,y)$,
governed by the three dimensionless groups $P$, $A$, and $r_0$, with the slope
$S$ as control parameter. The field is continuous everywhere and smooth away from
the line $x = 1$, across which the erosion source has a corner but no jump.

The physical quadrant $Q = \{x \geq 0,\; y \geq 0\}$ is forward-invariant. On the
edge $x = 0$ the speed derivative is $\mathrm{d}x/\mathrm{d}s = S y \geq 0$, which
points into $Q$, and on the edge $y = 0$ the load derivative is
$\mathrm{d}y/\mathrm{d}s = A\max(x^2-1,0) \geq 0$, which also points into $Q$;
hence a trajectory that starts with nonnegative speed and load remains in $Q$ for
all $s$. On a flat bed, $S = 0$, the speed equation~\eqref{eq:momentum_nondim}
decouples and reduces to $\mathrm{d}x/\mathrm{d}s = -P x^2$, whose solution from
any $x(0) = x_0 \geq 0$ is
\begin{equation}
x(s) \;=\; \frac{x_0}{1 + P\,x_0\,s} \;\xrightarrow[s\to\infty]{}\; 0 ,
\label{eq:flatbed_solution}
\end{equation}
so the speed decreases monotonically and falls below the erosion threshold
$x = 1$ in finite time. Once $x < 1$ the erosion source
in~\eqref{eq:sediment_nondim} vanishes for every closure that is zero below
threshold, and the load obeys $\mathrm{d}y/\mathrm{d}s = -r_0 y$ and decays
exponentially. The origin is therefore globally attracting on a flat bed, for any
disturbance and any admissible erosion law: a level cohesive bottom cannot sustain
a current, and this conclusion does not depend on the form of $E$.

The equilibria of~\eqref{eq:momentum_nondim}--\eqref{eq:sediment_nondim} are the
intersections of the nullclines. The speed nullcline $\mathrm{d}x/\mathrm{d}s = 0$
is the parabola $y = P x^2/S$, and the load nullcline $\mathrm{d}y/\mathrm{d}s =
0$ is $y = (A/r_0)\max(x^2-1,0)$, which coincides with the axis $y = 0$ for
$x \leq 1$ and rises as $(A/r_0)(x^2-1)$ for $x > 1$. The origin $(x,y) = (0,0)$
is always an equilibrium. In the erosion-on region $x > 1$, an interior
equilibrium $(x^\ast, y^\ast)$ satisfies both nullclines; eliminating $y$ through
$y = P x^2/S$ and substituting into $A(x^2-1) = r_0 y$ gives $A(x^2 - 1) =
(r_0 P/S)\,x^2$, whence
\begin{equation}
(x^\ast)^2 \;=\; \frac{A\,S}{A\,S - r_0\,P},
\qquad
y^\ast \;=\; \frac{P\,(x^\ast)^2}{S} .
\label{eq:interior_fp}
\end{equation}
This root is real, and exceeds unity so as to lie in the erosion-on region and be
physically admissible, if and only if the denominator is positive, $A S > r_0 P$.
That condition defines the critical slope
\begin{equation}
S_{\mathrm{crit}} \;=\; \frac{r_0\,P}{A} ,
\label{eq:scrit}
\end{equation}
below which the interior equilibrium is absent and above which it exists; at
$S = S_{\mathrm{crit}}$ the ignitive speed $x^\ast$ diverges, and as $S \to 1$ it
decreases toward a value just above the erosion threshold. The appearance of the
interior equilibrium at $S_{\mathrm{crit}}$ is a transcritical transition in the
control parameter $S$ \cite{strogatz2018}.

The linear stability of each equilibrium follows from the Jacobian of the vector
field, which in the erosion-on region $x > 1$ is
\begin{equation}
J(x,y) \;=\;
\begin{pmatrix}
\dfrac{\partial \dot x}{\partial x} & \dfrac{\partial \dot x}{\partial y} \\[8pt]
\dfrac{\partial \dot y}{\partial x} & \dfrac{\partial \dot y}{\partial y}
\end{pmatrix}
\;=\;
\begin{pmatrix}
-2\,P\,x & S \\[4pt]
2\,A\,x & -r_0
\end{pmatrix},
\label{eq:jacobian}
\end{equation}
with the lower-left entry replaced by zero in the erosion-off region $x < 1$. At
the interior equilibrium the trace and determinant are
\begin{equation}
\operatorname{tr} J(x^\ast,y^\ast) \;=\; -2\,P\,x^\ast - r_0 \;<\; 0,
\qquad
\det J(x^\ast,y^\ast) \;=\; 2\,x^\ast\big( r_0\,P - A\,S \big) .
\label{eq:trace_det}
\end{equation}
Wherever the interior equilibrium exists, $A S > r_0 P$, so the determinant is
negative. A planar equilibrium with negative determinant has real eigenvalues of
opposite sign and is a saddle \cite{strogatz2018}; the interior equilibrium is
therefore an unstable ignitive state, not a reachable steady flow. The
eigenvalues are the roots of $\lambda^2 - (\operatorname{tr} J)\lambda + \det J =
0$, one positive and one negative, and the eigenvector associated with the
negative eigenvalue spans the stable direction used below to construct the
separatrix.

The origin requires separate treatment because its linearization is degenerate.
Evaluating~\eqref{eq:jacobian} in the erosion-off region at $(0,0)$ gives the
matrix with rows $(0,\,S)$ and $(0,\,-r_0)$, whose eigenvalues are $0$ and
$-r_0$; the zero eigenvalue arises because the drag term $-P x^2$ has no linear
part at $x = 0$. The nonlinear flow nonetheless attracts the origin from within
$Q$: for any trajectory with $x < 1$ the load decays as $\mathrm{d}y/\mathrm{d}s =
-r_0 y$, and the speed obeys $\mathrm{d}x/\mathrm{d}s = S y - P x^2$ with a source
$S y$ that decays exponentially, so $x \to 0$ as well. The origin is thus an
attracting quiescent state, and the reduced model is bistable above
$S_{\mathrm{crit}}$ in the sense that trajectories tend either to the origin or to
unbounded growth, the two basins being separated by the stable manifold of the
saddle \cite{strogatz2018}. That manifold, the separatrix, is the exact boundary
between a self-limiting disturbance and an igniting one, and its intersection with
the load axis at $x = 0$ is the least lofted load, at zero initial speed, that
ignites; this critical lofted load, as a function of slope, is the safe-kick
envelope of the model.

\subsection{Numerical Implementation}

The reduced model is implemented as \texttt{siltout}, a small internal
\texttt{Python} package that solves no field equations and stores no simulation
archive, evaluating instead the vector field
of~\eqref{eq:momentum_nondim}--\eqref{eq:sediment_nondim}, its equilibria, its
Jacobian, and the diagnostics derived from them. All arithmetic is performed in
IEEE~754 double precision through \texttt{NumPy} \cite{harris2020}; the numerical
integration, root finding, and eigenvalue computations are provided by
\texttt{SciPy} \cite{virtanen2020}; and the figures are rendered with
\texttt{Matplotlib} \cite{hunter2007}. Each figure and report is produced by an
independent script that imports the package, so any result can be regenerated in
isolation.

The vector field is evaluated directly from its definition, with the erosion
source implemented as the sharp ramp $A\max(x^2-1,0)$ by default. For the
many-trajectory basin classifications an optional smooth surrogate replaces the
ramp by the softplus of width $\varepsilon$,
\begin{equation}
E_\varepsilon(x) \;=\; A\,\varepsilon\,\log\!\Big( 1 + e^{(x^2-1)/\varepsilon} \Big),
\qquad
E_\varepsilon(x) \;\xrightarrow[\varepsilon\to 0^+]{}\; A\,\max(x^2-1,0),
\label{eq:softplus}
\end{equation}
which removes the corner at $x = 1$ and the attendant step-size penalty of an
adaptive integrator while converging to the sharp ramp as $\varepsilon \to 0^+$;
the sharp closure is used for every figure in which the separatrix and the
classified samples must coincide. The Jacobian~\eqref{eq:jacobian} is assembled
analytically and, in the verification stage, is checked against a second-order
central finite-difference Jacobian formed from the vector field.

Time integration uses the implicit, L-stable Radau method of order five, a member
of the family of implicit Runge--Kutta schemes for stiff systems
\cite{hairerwanner1996}, applied with relative tolerance $10^{-10}$ and absolute
tolerance $10^{-12}$. The implicit scheme is chosen because the reduced system is
mildly stiff near ignition, where the two Jacobian eigenvalues differ by an order
of magnitude, so that an explicit method would be step-size limited by the fast
stable mode; the L-stability of the Radau method integrates the system without
that restriction. As an independent check, selected trajectories are re-integrated
with the explicit eighth-order Dormand--Prince method \cite{dormandprince1980} at
the same tolerances, and the two solutions are required to agree. For the basin
classifications, which require many short integrations rather than accurate
trajectories, a lean adaptive Runge--Kutta pass is used with two terminal events:
a runaway event that stops the integration when the speed exceeds a fixed multiple
of the ignitive speed, signaling ignition, and a settling event that stops the
integration when the speed and load both fall below a small threshold, signaling
decay. The softplus surrogate~\eqref{eq:softplus} is enabled for this pass alone.

The interior equilibrium is evaluated from its closed form~\eqref{eq:interior_fp}
and, independently, located as the unique root in $x > 1$ of $h(x) = A(x^2-1) -
(r_0 P/S)x^2$ by Brent's method on the bracket $[1^+,\,x_{\max}]$, the two
determinations being required to agree. The stable manifold of the saddle is
constructed by forming the Jacobian~\eqref{eq:jacobian} at $(x^\ast,y^\ast)$,
computing its eigenvalues and eigenvectors, selecting the eigenvector
$\mathbf{v}_-$ of the negative eigenvalue as the stable direction, seeding two
points $(x^\ast,y^\ast) \pm \delta\,\mathbf{v}_-$ at a small offset $\delta$ on
either side of the equilibrium, and integrating the vector field backward in $s$
from each seed; the union of the two backward orbits, restricted to $Q$, is the
separatrix \cite{strogatz2018}. The critical lofted load $y_0^{\mathrm{crit}}(S)$
that separates decay from ignition at a fixed slope is located by bisection on the
classifier outcome over the initial load at zero initial speed, and the
operational critical slope is located by bisection on the classifier outcome over
the slope at a fixed bounded disturbance.

\subsection{Data Analysis}

Five diagnostic analyses are applied to the reduced model, each implemented as an
independent script that regenerates its figure and its tabulated output. The
fixed-point analysis evaluates the closed-form equilibrium~\eqref{eq:interior_fp}
over a set of supercritical slopes, forms the Jacobian~\eqref{eq:jacobian} at each
equilibrium and at the origin, computes the eigenvalues, classifies each
equilibrium from their signs, and records the residual of the vector field at each
equilibrium as a confirmation that it has been located to machine precision; the
phase portrait is drawn at a representative supercritical slope with the
nullclines, the equilibria, the separatrix, and two trajectories that straddle it.
The bifurcation analysis evaluates the ignitive speed $x^\ast$ across the slope
and records the branch and its onset at $S_{\mathrm{crit}}$, and the scaling study
varies the critical shear $\tau_c$ and the cloud thickness $h_0$ in turn to record
the critical bed angle as a function of each. The envelope analysis constructs the
separatrix and the critical lofted load $y_0^{\mathrm{crit}}(S)$, and validates
the basins by classifying a grid of launched trajectories against the separatrix.
The verification analysis performs four independent checks: a tolerance
self-convergence study of the terminal state against a tightly converged
reference; a cross-integrator comparison of the Radau and Dormand--Prince
solutions along a common trajectory; a fixed-point residual study; and a
comparison of the analytic and finite-difference Jacobian eigenvalues. The
robustness analysis tests the flat-bed self-limiting result for large
disturbances, compares the operational and closed-form critical slopes over a
range of the near-bed ratio $r_0$, records the operational critical slope as a
function of the regularization width $\varepsilon$, and records it for the family
of erosion exponents $A\max(x^2-1,0)^p$ that generalizes the linear law.

\section{Results}

The reduced model was evaluated at a representative parameter set for a strong fin
disturbance over fine cohesive silt in seawater, listed with the derived
dimensionless quantities in Table~\ref{tab:parameters}. All values reported below
are taken directly from the diagnostic output and are quoted to three decimal
places; quantities are dimensionless unless a unit is stated. The submerged
specific gravity is $R = 1.585$, the erosion-threshold speed is $U_c = 0.099~
\mathrm{m\,s^{-1}}$, the erosion magnitude is $\alpha = 1.415 \times 10^{-4}$, and
the settling timescale is $t_{\mathrm{settle}} = h_0/w_s = 375~\mathrm{s}$, about
$6.2$ minutes. The three dimensionless groups that close the model are the
drag-settling number $P = 1.235$, the erosion-strength number $A = 8.356$, and the
near-bed concentration ratio $r_0 = 2$, which give the critical slope
$S_{\mathrm{crit}} = r_0 P/A = 0.296$ and the critical bed angle
$\theta_{\mathrm{crit}} = \arcsin S_{\mathrm{crit}} = 17.189^\circ$.

\begin{table}[H]
\centering
\caption{Physical parameters of the representative case and the derived
dimensionless quantities. The upper block lists the dimensional inputs; the lower
block lists the derived scales and the three dimensionless groups $P$, $A$, and
$r_0$ that close the reduced model, together with the critical slope
$S_{\mathrm{crit}} = r_0 P/A$ and the critical bed angle.}
\label{tab:parameters}
\begin{tabular}{lll}
\toprule
Quantity & Symbol & Value \\
\midrule
Clear-water density          & $\rho_0$ & $1025~\mathrm{kg\,m^{-3}}$ \\
Grain density                & $\rho_s$ & $2650~\mathrm{kg\,m^{-3}}$ \\
Gravitational acceleration   & $g$      & $9.810~\mathrm{m\,s^{-2}}$ \\
Settling velocity            & $w_s$    & $8.000 \times 10^{-4}~\mathrm{m\,s^{-1}}$ \\
Critical shear stress        & $\tau_c$ & $0.100~\mathrm{Pa}$ \\
Drag coefficient             & $C_d$    & $1.000 \times 10^{-2}$ \\
Cloud thickness              & $h_0$    & $0.300~\mathrm{m}$ \\
Erosion-rate parameter       & $M$      & $3.000 \times 10^{-4}~\mathrm{kg\,m^{-2}\,s^{-1}}$ \\
Near-bed concentration ratio & $r_0$    & $2.000$ \\
\midrule
Submerged specific gravity   & $R$      & $1.585$ \\
Kinematic critical shear     & $\tau_c/\rho_0$ & $9.756 \times 10^{-5}~\mathrm{m^2\,s^{-2}}$ \\
Erosion-threshold speed      & $U_c$    & $0.099~\mathrm{m\,s^{-1}}$ \\
Erosion magnitude            & $\alpha$ & $1.415 \times 10^{-4}$ \\
Settling timescale           & $t_{\mathrm{settle}}$ & $375.000~\mathrm{s}$ \\
Drag-settling number         & $P$      & $1.235$ \\
Erosion-strength number      & $A$      & $8.356$ \\
Critical slope               & $S_{\mathrm{crit}}$ & $0.296$ \\
Critical bed angle           & $\theta_{\mathrm{crit}}$ & $17.189^\circ$ \\
\bottomrule
\end{tabular}
\end{table}

The equilibrium structure is reported in Table~\ref{tab:fixedpoints}. The quiescent
state at the origin is present at every slope, with the degenerate eigenvalue pair
$(0,\,-2)$, the second eigenvalue being $-r_0$; it is an attracting quiescent
state as established in the derivation. No interior equilibrium exists at $S = 0$
or at $S = S_{\mathrm{crit}} = 0.296$, consistent with the onset of the branch
exactly at the critical slope. At each supercritical slope the interior
equilibrium is a saddle, its eigenvalues comprising one negative and one positive
real value: $(-22.256,\, +0.027)$ at $S = 0.300$, $(-6.663,\, +0.802)$ at
$S = 0.500$, $(-6.695,\, +1.586)$ at $S = 0.800$, $(-6.820,\, +1.807)$ at
$S = 0.900$, and $(-6.958,\, +2.016)$ at $S = 1.000$. The closed-form
equilibrium~\eqref{eq:interior_fp} agrees with the independent root determination
to between $4.441 \times 10^{-16}$ and $4.796 \times 10^{-14}$ in the speed
coordinate, and the residual of the vector field at each equilibrium is between
$0$ and $2.665 \times 10^{-15}$, at the level of machine precision. The ignitive
speed decreases from $x^\ast = 8.192$ at $S = 0.300$, just above the critical
slope where the branch diverges, to $x^\ast = 1.191$ at $S = 1.000$, so the
ignitive state on a steep bed is a slow current poised at the edge of erosion
rather than a fast one.

\begin{table}[H]
\centering
\caption{Interior equilibrium and its linear stability at a sequence of
supercritical bed slopes. The ignitive speed $x^\ast$ and load $y^\ast$ are the
closed form~\eqref{eq:interior_fp}; $\lambda_1$ and $\lambda_2$ are the real
eigenvalues of the Jacobian~\eqref{eq:jacobian}; and the residual is the max-norm
of the vector field at the equilibrium. The equilibrium is a saddle at every
slope, and no interior equilibrium exists below $S_{\mathrm{crit}} = 0.296$, where
the model self-limits for every disturbance.}
\label{tab:fixedpoints}
\begin{tabular}{lccccc}
\toprule
$S$ & $x^\ast$ & $y^\ast$ & $\lambda_1$ & $\lambda_2$ & Residual \\
\midrule
0.300 & 8.192 & 276.186 & $-22.256$ & $+0.027$ & $0 \times 10^{0}$ \\
0.500 & 1.564 & 6.038   & $-6.663$  & $+0.802$ & $1.776 \times 10^{-15}$ \\
0.800 & 1.259 & 2.447   & $-6.695$  & $+1.586$ & $1.776 \times 10^{-15}$ \\
0.900 & 1.220 & 2.043   & $-6.820$  & $+1.807$ & $2.665 \times 10^{-15}$ \\
1.000 & 1.191 & 1.753   & $-6.958$  & $+2.016$ & $4.441 \times 10^{-16}$ \\
\bottomrule
\end{tabular}
\end{table}

\begin{figure}[H]
\centering
\includegraphics[width=0.62\linewidth]{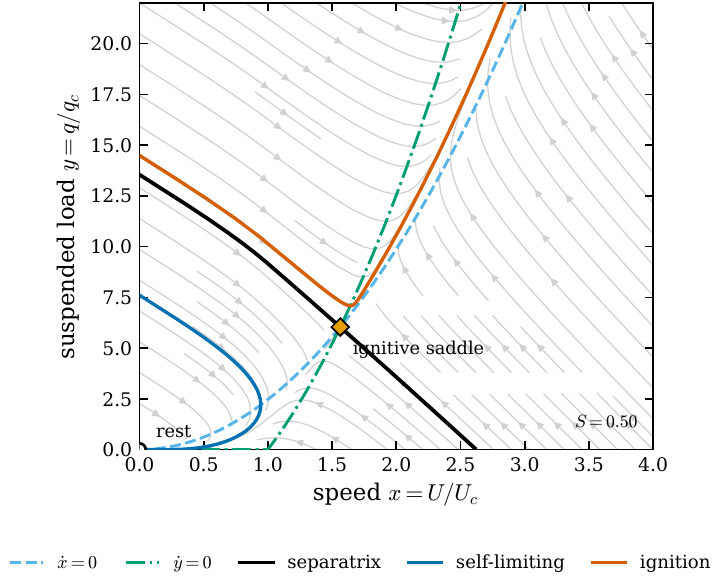}
\caption{Phase portrait of the reduced model at the supercritical slope
$S = 0.500$. The dashed curve is the speed nullcline $\dot x = 0$ and the
dot-dashed curve is the load nullcline $\dot y = 0$; their intersection in the
erosion-on region is the ignitive saddle, shown with the quiescent rest state at
the origin. The solid black curve is the separatrix, the stable manifold of the
saddle. The blue trajectory is launched below the separatrix and self-limits; the
vermillion trajectory is launched above it and ignites. Gray streamlines show the
vector field.}
\label{fig:phase}
\end{figure}

The phase portrait at the supercritical slope $S = 0.500$, a bed angle of
$30^\circ$, is shown in Figure~\ref{fig:phase}. The speed nullcline and the load
nullcline intersect at the ignitive saddle $(x^\ast,y^\ast) = (1.564,\, 6.038)$,
marked together with the quiescent rest state at the origin. The separatrix, the
stable manifold of the saddle computed by backward integration, divides the
physical quadrant into the basin of the rest state and the basin of unbounded
growth. A trajectory launched from a lofted load below the separatrix rises
briefly in speed, sheds its load to deposition, and returns to rest, whereas a
trajectory launched from a lofted load above the separatrix accelerates past the
erosion threshold $x = 1$ and runs away; these are the self-limiting and the
igniting outcomes the model is built to distinguish.

\begin{figure}[H]
\centering
\includegraphics[width=\linewidth]{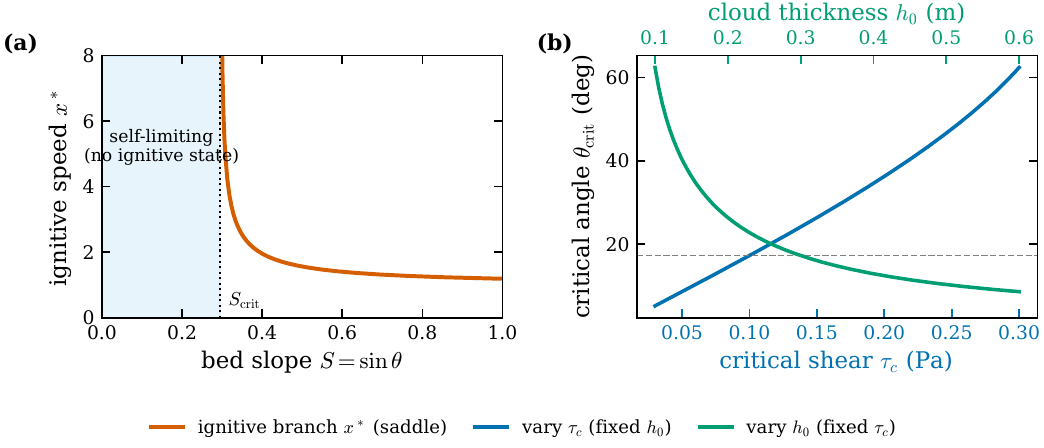}
\caption{Ignition bifurcation and critical-slope scaling. (a) Ignitive speed
$x^\ast$ against bed slope $S$, from equation~\eqref{eq:interior_fp}; the branch
appears at $S_{\mathrm{crit}} = r_0 P/A$ and is a saddle wherever it exists, and
the shaded region below $S_{\mathrm{crit}}$ is the range in which the model
self-limits for every disturbance. (b) Critical bed angle $\theta_{\mathrm{crit}}$
as the critical shear $\tau_c$ (lower axis) and the cloud thickness $h_0$ (upper
axis) are varied in turn, with the remaining parameters fixed; the dashed
reference marks the base value $17.189^\circ$.}
\label{fig:bifurcation}
\end{figure}

The dependence of the ignitive equilibrium on the bed slope is the transcritical
appearance shown in Figure~\ref{fig:bifurcation}(a): the branch of ignitive speeds
is absent below $S_{\mathrm{crit}} = 0.296$, appears at the critical slope,
diverges as the slope approaches it from above, and decreases monotonically toward
the erosion threshold as the slope steepens. The branch is a saddle at every slope
at which it exists, so it is the ignition threshold rather than a reachable steady
state, and the shaded region below the critical slope is the parameter range in
which no self-sustaining current is possible for any disturbance.
Figure~\ref{fig:bifurcation}(b) reports the critical bed angle as the critical
shear stress and the cloud thickness are varied in turn: a more resistant bed of
higher $\tau_c$ requires a steeper slope to ignite, so the critical angle rises
with $\tau_c$, while a thicker cloud carries more buoyant sediment per unit bed
area and ignites on a gentler slope, so the critical angle falls with $h_0$. The
critical angle passes through its base value of $17.189^\circ$ at the
representative parameters and spans roughly one to six tens of degrees over the
ranges examined, so the threshold is sensitive to the bed and disturbance
properties but remains a moderate slope for physically reasonable values.

\begin{figure}[H]
\centering
\includegraphics[width=\linewidth]{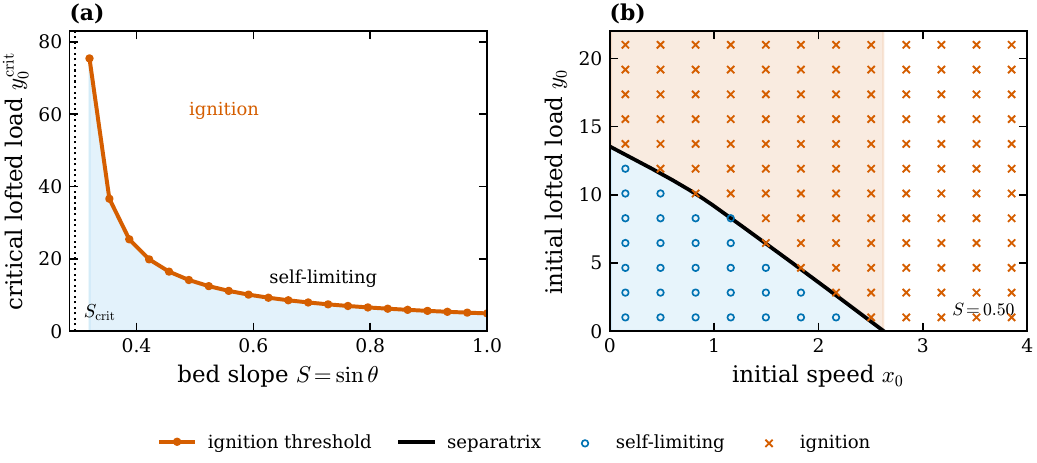}
\caption{Safe-kick envelope and basins of attraction. (a) Critical lofted load
$y_0^{\mathrm{crit}}$, the least initial load at zero initial speed that ignites,
against bed slope $S$; it diverges at $S_{\mathrm{crit}}$ and falls as the slope
steepens, and the shaded region below the envelope is the self-limiting regime.
(b) Basins of attraction in the plane of initial speed $x_0$ and initial load
$y_0$ at $S = 0.500$: the separatrix (solid) divides the plane, and independently
integrated samples (open circles, self-limiting; crosses, ignition) confirm the
boundary.}
\label{fig:envelope}
\end{figure}

Above the critical slope the outcome of a disturbance depends on its magnitude
relative to the separatrix, quantified by the safe-kick envelope of
Figure~\ref{fig:envelope}(a). The critical lofted load $y_0^{\mathrm{crit}}$, the
least initial load at zero initial speed that ignites, diverges as the slope
approaches $S_{\mathrm{crit}}$ from above and falls monotonically as the slope
steepens: it takes the values $38.492$ at $S = 0.350$ (a bed angle of
$20.487^\circ$), $13.539$ at $S = 0.500$ ($30.000^\circ$), $8.693$ at $S = 0.650$
($40.542^\circ$), $6.482$ at $S = 0.800$ ($53.130^\circ$), and $4.869$ at
$S = 1.000$ ($90.000^\circ$), so a steeper bed is carried into a self-sustaining
silt-out by a weaker disturbance. The region below the envelope is the
self-limiting regime and the region above it the igniting regime.
Figure~\ref{fig:envelope}(b) shows the basins of attraction in the plane of
initial speed and initial load at $S = 0.500$: the separatrix divides the plane,
and a grid of independently integrated trajectories, each classified by direct
time integration, falls on the correct side of the separatrix at every point,
including near the boundary itself, which confirms that the stable manifold
computed by backward integration is the exact basin boundary.

\begin{figure}[H]
\centering
\includegraphics[width=\linewidth]{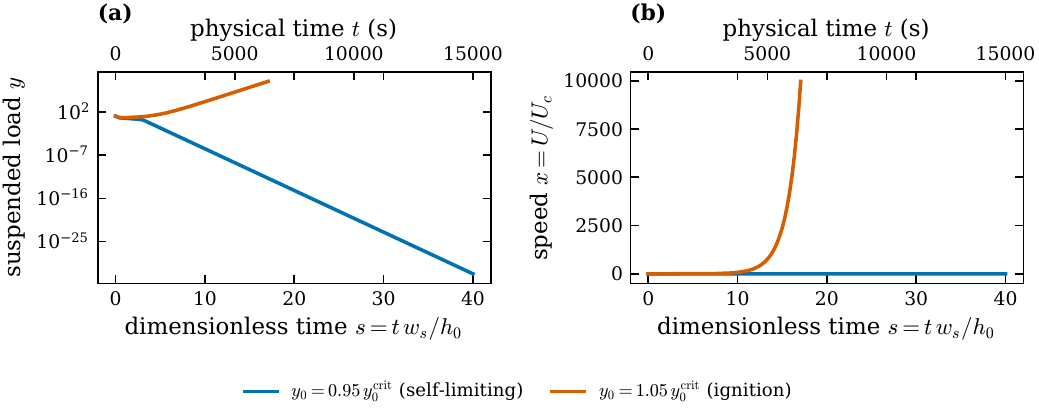}
\caption{Representative time series straddling the ignition threshold at
$S = 0.500$. (a) Suspended load $y$ on a logarithmic ordinate for disturbances at
$0.95\,y_0^{\mathrm{crit}}$ (self-limiting) and $1.05\,y_0^{\mathrm{crit}}$
(ignition). (b) Speed $x$ for the same two disturbances. The upper axis reports
physical time through the settling timescale $t_{\mathrm{settle}} =
375~\mathrm{s}$.}
\label{fig:timeseries}
\end{figure}

The evolution in time of two disturbances straddling the threshold at $S = 0.500$
is shown in Figure~\ref{fig:timeseries}. A disturbance at ninety-five percent of
the critical load decays, its load falling exponentially on the settling timescale
and its speed relaxing toward zero, while a disturbance at one hundred and five
percent of the critical load ignites, its load and speed growing without bound in
the minimal model with the runaway reached within a few tens of dimensionless time
units. The dimensionless time is referred to the settling timescale
$t_{\mathrm{settle}} = 375~\mathrm{s}$ on the secondary axis, which places the
exponential clearing of a self-limiting cloud at a few minutes, consistent with
the observed persistence of a diver-induced silt-out that does not ignite.

\begin{figure}[H]
\centering
\includegraphics[width=\linewidth]{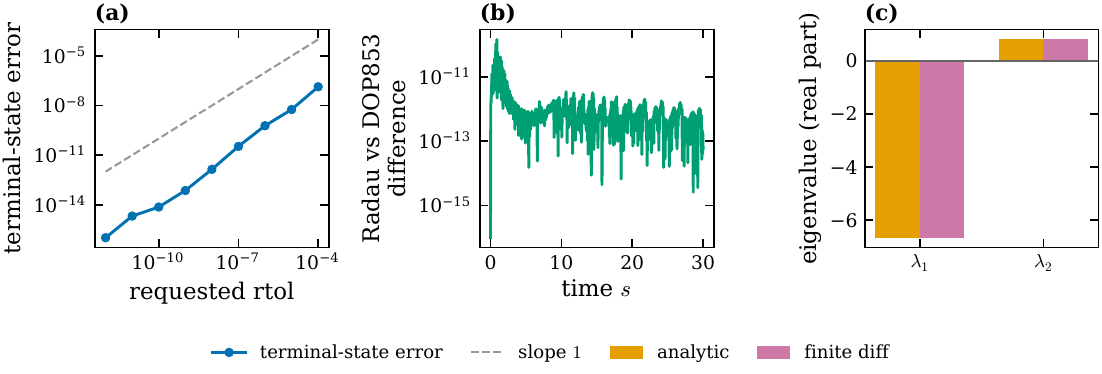}
\caption{Numerical verification of the reduced-model solver. (a) Terminal-state
error against requested relative tolerance for a self-limiting trajectory, with a
reference slope of unity. (b) Difference between the implicit Radau and explicit
Dormand--Prince solutions along a common trajectory. (c) Real parts of the
analytic and finite-difference Jacobian eigenvalues at the interior equilibrium
for $S = 0.500$.}
\label{fig:verification}
\end{figure}

The numerical reliability of the solver is established by the four checks
summarized in Table~\ref{tab:verification} and shown in
Figure~\ref{fig:verification}. The tolerance self-convergence study, in
Figure~\ref{fig:verification}(a), integrates a self-limiting trajectory at
relative tolerances from $10^{-4}$ to $10^{-12}$ and measures the terminal-state
error against a reference converged at relative tolerance $10^{-13}$; the error
falls monotonically from $1.360 \times 10^{-7}$ at $10^{-4}$ through
$6.012 \times 10^{-10}$ at $10^{-6}$, $1.375 \times 10^{-12}$ at $10^{-8}$, and
$7.185 \times 10^{-15}$ at $10^{-10}$, to $2.776 \times 10^{-17}$ at $10^{-12}$,
tracking the requested tolerance down to the double-precision floor. The
cross-integrator study, in Figure~\ref{fig:verification}(b), finds a maximum
difference of $1.392 \times 10^{-10}$ between the Radau and Dormand--Prince
solutions along a common trajectory. The fixed-point residual study confirms that
the closed-form equilibria annul the vector field to $1.776 \times 10^{-15}$ at
$S = 0.800$, $2.665 \times 10^{-15}$ at $S = 0.900$, and $4.441 \times 10^{-16}$
at $S = 1.000$. The Jacobian study, in Figure~\ref{fig:verification}(c), finds
that the eigenvalues of the analytic Jacobian at the $S = 0.500$ equilibrium,
$-6.663$ and $0.802$, agree with those of the finite-difference Jacobian to a
maximum absolute difference of $1.348 \times 10^{-10}$.

\begin{table}[H]
\centering
\caption{Numerical verification summary. The tolerance self-convergence lists the
terminal-state error at the loosest and tightest tolerances; the cross-integrator
agreement is the maximum trajectory difference between the Radau and
Dormand--Prince solutions; the fixed-point residual is the range of vector-field
max-norms at the supercritical equilibria; and the Jacobian agreement is the
maximum absolute difference between the analytic and finite-difference eigenvalues
at $S = 0.500$.}
\label{tab:verification}
\begin{tabular}{ll}
\toprule
Check & Result \\
\midrule
Terminal error at $\mathrm{rtol} = 10^{-4}$   & $1.360 \times 10^{-7}$ \\
Terminal error at $\mathrm{rtol} = 10^{-12}$  & $2.776 \times 10^{-17}$ \\
Radau versus Dormand--Prince (max difference) & $1.392 \times 10^{-10}$ \\
Fixed-point residual (supercritical range)    & $4.441 \times 10^{-16}$ to $2.665 \times 10^{-15}$ \\
Analytic versus finite-difference eigenvalues & $1.348 \times 10^{-10}$ \\
\bottomrule
\end{tabular}
\end{table}

\begin{figure}[H]
\centering
\includegraphics[width=\linewidth]{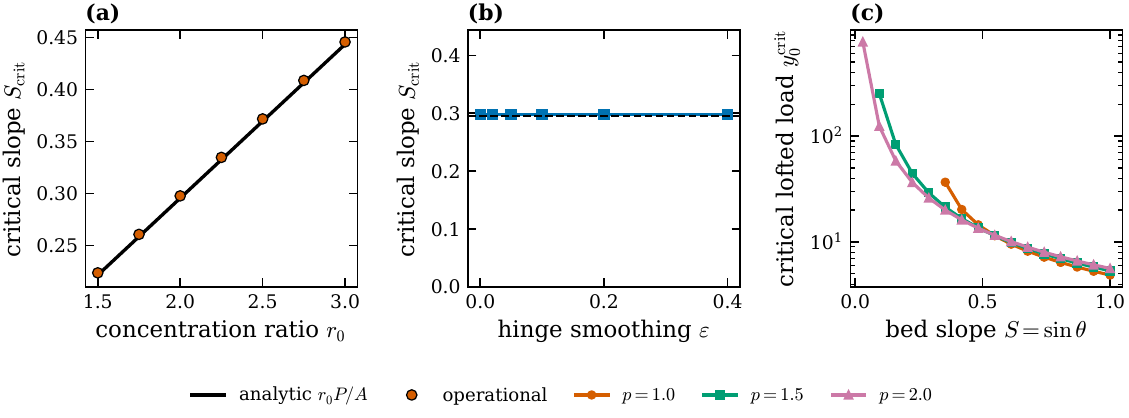}
\caption{Robustness of the ignition threshold. (a) Critical slope against the
near-bed concentration ratio $r_0$: the analytic line $r_0 P/A$ and the
operational values from integration coincide. (b) Operational critical slope
against the softplus regularization width $\varepsilon$, which is flat. (c)
Safe-kick envelope for erosion exponents $p = 1.0$, $1.5$, and $2.0$: each has a
finite positive threshold, the linear closure gives the highest threshold, and
superlinear erosion lowers it.}
\label{fig:robustness}
\end{figure}

The robustness of the ignition threshold to the erosion closure is reported in
Table~\ref{tab:robustness} and shown in Figure~\ref{fig:robustness}. The flat-bed
result holds by construction and independently of the closure, as established
through the explicit solution~\eqref{eq:flatbed_solution}, and is confirmed
numerically: at zero slope the model self-limits for lofted loads as large as
$50$, $200$, and $1000$. Above zero slope the operational critical slope obtained
by integration, $0.298$, matches the closed-form value $0.296$ to within $0.002$,
the small excess arising from the finite disturbance cap used to detect ignition
near the divergence of the envelope. Figure~\ref{fig:robustness}(a) shows that the
operational critical slope scales linearly with the near-bed concentration ratio
$r_0$ and coincides with the closed-form line $r_0 P/A$: the operational values
are $0.224$, $0.298$, $0.372$, and $0.446$ at $r_0 = 1.5$, $2.0$, $2.5$, and
$3.0$, against the closed-form values $0.222$, $0.296$, $0.369$, and $0.443$.
Figure~\ref{fig:robustness}(b) shows that the operational critical slope is
invariant to the softplus width, remaining at $0.298$ as $\varepsilon$ ranges over
$0.0$, $0.05$, $0.10$, $0.20$, and $0.40$, so the threshold is not an artifact of
the corner in the sharp closure. Figure~\ref{fig:robustness}(c) shows the
safe-kick envelope for three erosion exponents; a finite positive threshold exists
for each, the envelopes converge at steep slopes, and the operational critical
slope falls from $0.298$ for the linear closure through $0.117$ at $p = 1.25$ and
$0.066$ at $p = 1.50$ to $0.035$ for the quadratic closure. The linear closure
yields the highest critical slope and is therefore the conservative choice, while
superlinear erosion lowers the slope at which a bounded disturbance ignites without
removing the threshold.

\begin{table}[H]
\centering
\caption{Robustness summary of the ignition threshold. The linear-closure
operational critical slope is compared with the closed-form value; the scaling
with the near-bed ratio $r_0$ and the invariance to the regularization width
$\varepsilon$ are quoted at their endpoints; and the operational critical slope is
quoted for four erosion exponents $p$ in the family $A\max(x^2-1,0)^p$. The linear
closure $p = 1$ is the conservative choice.}
\label{tab:robustness}
\begin{tabular}{lll}
\toprule
Check & Setting & Operational $S_{\mathrm{crit}}$ \\
\midrule
Closed form (linear)  & $r_0 P/A$           & $0.296$ \\
Integration (linear)  & bounded disturbance & $0.298$ \\
Concentration ratio   & $r_0 = 1.5$         & $0.224$ \\
                      & $r_0 = 3.0$         & $0.446$ \\
Regularization width  & $\varepsilon = 0.0$ & $0.298$ \\
                      & $\varepsilon = 0.4$ & $0.298$ \\
Erosion exponent      & $p = 1.00$          & $0.298$ \\
                      & $p = 1.25$          & $0.117$ \\
                      & $p = 1.50$          & $0.066$ \\
                      & $p = 2.00$          & $0.035$ \\
\bottomrule
\end{tabular}
\end{table}

\section{Discussion}

The reduced model reproduces, from a two-variable planar vector field derived at
the scale of a diver disturbance, the ignition structure established for erosive
turbidity currents \cite{parker1982}. The interior equilibrium is a saddle
wherever it exists, so it is an unstable ignitive state rather than a reachable
steady flow; disturbances below its stable manifold decay to the quiescent state,
and disturbances above it accelerate past the erosion threshold and run away. This
is the same dichotomy, between a self-limiting flow that dies out and a
self-sustaining flow that ignites, that the phase-plane analysis of autosuspension
delivers for turbidity currents \cite{pantin1979,bagnold1962}, now obtained in
closed form for the diver-scale problem with the bed slope as the explicit control
parameter. The correspondence is structural rather than incidental, since the
reduced vector field descends by an explicit and justified reduction from the
layer-averaged balances that underlie the three-equation turbidity-current model
\cite{pfp1986,meiburg2010}, and the closed-form critical slope $S_{\mathrm{crit}}
= r_0 P/A$ is the diver-scale expression of the autosuspension threshold.

The physical reading is direct. A flat cohesive bottom cannot sustain a silt-out,
a conclusion that is closure-independent because it follows from the explicit
solution~\eqref{eq:flatbed_solution} of the speed equation at zero slope rather
than from any property of the erosion law; a disturbance of any magnitude settles
back on the settling timescale, which for the representative cloud is about $6.2$
minutes, matching the operational experience that a silt-out over level ground
clears on its own if it is not disturbed further. A self-sustaining current
requires a bed steeper than the critical angle, close to $17^\circ$ for the
representative silt, a moderate slope of the kind encountered on reef faces, sand
chutes, and drop-off approaches rather than a wall. Above that angle the outcome
depends on the disturbance magnitude relative to the separatrix, and a steeper bed
is crossed into ignition by a weaker disturbance, so the hazard grows with slope in
two distinct ways: through the appearance of the threshold at $S_{\mathrm{crit}}$
and through the fall of the safe-kick envelope above it. These readings identify
the bed slope as the controlling parameter and the resuspended load as the
disturbance measure, which is the qualitative guidance the model is positioned to
provide.

The robustness study delimits how far these conclusions depend on the erosion
closure, and its results are reported without overstatement. The existence of a
finite positive critical slope is robust: it holds for the linear excess-shear
law, for its smooth regularization, and for the superlinear closures, and the
flat-bed self-limiting result holds for every admissible closure by an argument
that does not reference the closure at all. The value of the critical slope, by
contrast, is closure-dependent, and the study frames the linear closure as the
conservative choice because it yields the highest threshold, with superlinear
erosion lowering the slope at which a bounded disturbance ignites. The linear
scaling with the near-bed ratio and the invariance to the regularization width
confirm that the closed-form threshold is recovered by integration and is not an
artifact of the corner in the sharp closure. The honest summary is that the model
predicts the existence and the controlling parameter of the threshold firmly, and
its numerical value only to within the latitude of the erosion law, a latitude
that the experimental literature on autosuspension has itself emphasized
\cite{southard1981,garciaparker1991}.

Several assumptions bound the scope of the model and are stated here rather than
left to inference. The fixed-thickness slab of the reduction neglects the
entrainment of ambient water and the growth of the layer that accompany a
developing current; because entrainment dilutes the suspension and reduces the
deposition flux per unit bed area, the fixed-thickness treatment overstates
deposition and is conservative in the direction of a higher threshold, which
justifies its use for a bounding estimate. The minimal model carries no saturation
of the suspension, so an ignited current grows without bound rather than settling
to the finite catastrophic state that the full theory predicts \cite{parker1982};
ignition is accordingly read from the crossing of the threshold rather than from
the long-time asymptotics, and a saturation closure, through hindered settling or a
capacity limit, would restore the stable catastrophic state and convert the
picture into a complete saddle-node structure with a second attracting equilibrium
at high load \cite{strogatz2018}. The single drag coefficient identifies the skin
friction that mobilizes sediment with the form drag that retards the layer,
whereas the two differ in a resolved boundary layer; this is a standard
simplification of layer-averaged modeling, itself the subject of continuing
assessment \cite{hu2015}, and it shifts the value of the threshold through the
erosion-threshold speed without affecting its existence. The erosion and
deposition closures are empirical, and the settling velocity is held constant, so
the model does not represent size sorting or hindered settling. Most importantly
for any operational reading, the disturbance in the model is the dimensionless
lofted load, and its mapping to a physical fin thrust is an order-of-magnitude
correspondence rather than a calibration; the model reports dimensionless
structure and the identity of the controlling parameter, and it is not a
quantitative safety guideline. The reduced formulation avoids the
turbulence-closure objection that attends a diver-scale field solver
\cite{meiburg2015}, since it resolves no turbulence and makes its idealizations
explicit in the derivation, but the empirical status of its closures is the price
of that transparency. Placing the model on an empirical footing would require
controlled resuspension experiments over sloping cohesive beds against which the
critical slope and the safe-kick envelope could be calibrated, and would admit the
entrainment and saturation closures as the natural next terms, which is the
direction in which the present idealized treatment extends.

\section{Conclusions}

This work derived a reduced-order dynamical model for the ignition of
diver-induced cohesive silt-out by reducing the layer-averaged balances of fluid
mass, sediment mass, and momentum, in the weakly entraining limit of a slow
near-bed cloud, to a temporal slab of fixed thickness, obtaining an autonomous
planar vector field in a dimensionless near-bed speed and a dimensionless
suspended load governed by three dimensionless groups with the bed slope as
control parameter, and it established in closed form that this field carries a
quiescent equilibrium at the origin, globally attracting on a flat bed for every
admissible erosion law, together with, above a critical bed slope
$S_{\mathrm{crit}} = r_0 P/A = 0.296$, an interior equilibrium whose negative
Jacobian determinant marks it as a saddle, so that the model reproduces the
classical ignition structure of erosive gravity currents in which disturbances
below the saddle's stable manifold self-limit and disturbances above it ignite,
with the stable manifold serving as the safe-kick envelope, the critical angle
falling near $17.189^\circ$ for representative silt, the closed-form structure
confirmed against time integration to machine precision, against a second
integrator to ten decimal places, and against a finite-difference Jacobian to
better than $10^{-9}$, and the threshold shown to be robust in existence and
controlling parameter though closure-dependent in value, so that the model
delivers a transparent, verified, and openly reproducible account of the
slope-controlled transition between a silt-out that clears on its own and one that
sustains itself, while leaving the mapping from the dimensionless disturbance to a
physical thrust, and the calibration of the threshold against controlled
resuspension experiments, as the natural next steps.

\section*{Acknowledgements}
The authors used Claude Opus~4.8 (Anthropic, PBC) solely as a writing-assistance
tool to refine English vocabulary and grammar during the preparation of this
manuscript. All scientific content, the derivation, the numerical implementation,
the analyses, the interpretations, the conclusions, and any remaining linguistic
imperfections are the sole responsibility of the authors.

\section*{Funding}
This research was supported by the Directorate of Research and Innovation, Bandung Institute of Technology, under the ITB 3P Research Program 2026 (Talenta Unggul Scheme), Project ID: DRI.PN-6-64-2026.

\section*{Data Availability}
Python source code and the analysis scripts that reproduce every
figure, table, and diagnostic report in this study are available on GitHub at
\url{https://github.com/sandyherho/scuba_siltout_reduced}. All supplementary
outputs, comprising the diagnostic reports, the tabulated data behind each figure,
and the figures in vector and raster formats, are permanently archived on the Open
Science Framework at \url{https://doi.org/10.17605/OSF.IO/ER8K9}. The source
code, the analysis scripts, and the archived outputs are released under the MIT license.

\end{document}